\documentclass[twocolumn]{emulateapj}
\usepackage[utf8]{inputenc}
\graphicspath{{figs/}} %
\usepackage[colorlinks,linkcolor=blue,anchorcolor=blue,citecolor=blue]{hyperref}
\usepackage[hyperref]{backref}
\usepackage{CJK}

\usepackage{graphicx} 
\usepackage{txfonts}
\usepackage{natbib}
\begin{document}
\begin{CJK*}{UTF8}{gbsn}

\title{Inside a Beehive: the Multiple Merging Processes in the Galaxy Cluster Abell 2142}


\author{Ang Liu\altaffilmark{1,2,3,4}(刘昂), Heng Yu\altaffilmark{1}(余恒), Antonaldo
Diaferio\altaffilmark{5,6}, Paolo Tozzi\altaffilmark{2,1}, Ho Seong Hwang\altaffilmark{7}, Keiichi Umetsu\altaffilmark{8}, Nobuhiro Okabe\altaffilmark{9,10,11}, Li-Lan Yang\altaffilmark{1,12}(杨里岚)}

\altaffiltext{1}{Department of Astronomy, Beijing Normal University, Beijing, China 100875;
                 yuheng@bnu.edu.cn, liuang@arcetri.astro.it}
\altaffiltext{2}{INAF Osservatorio Astrofisico di Arcetri, Largo E. Fermi, I-50122 Firenze, Italy}
\altaffiltext{3}{Department of Physics, Sapienza University of Rome, I-00185, Rome, Italy}
\altaffiltext{4}{Department of Physics, University of Rome Tor Vergata, I-00133, Rome, Italy}
\altaffiltext{5}{Dipartimento di Fisica, Universit\`a di Torino, Via P. Giuria 1, I-10125 Torino, Italy}
\altaffiltext{6}{Istituto Nazionale di Fisica Nucleare (INFN), Sezione di Torino, Via P. Giuria 1, I-10125 Torino, Italy}
\altaffiltext{7}{Quantum Universe Center, Korea Institute for Advanced Study, 85 Hoegiro, Dongdaemun-gu, Seoul 02455, Korea}
\altaffiltext{8}{Institute of Astronomy and Astrophysics, Academia Sinica, P.O. Box 23-141, Taipei 10617, Taiwan}
\altaffiltext{9}{Department of Physical Science, Hiroshima University, 1-3-1, Kagamiyama, Higashi-Hiroshima, Hiroshima 739-8526, Japan}
\altaffiltext{10}{Hiroshima Astrophysical Science Center, Hiroshima University, 1-3-1, Kagamiyama, Higashi-Hiroshima, Hiroshima 739-8526, Japan}
\altaffiltext{11}{Core Research for Energetic Universe, Hiroshima University, 1-3-1, Kagamiyama, Higashi-Hiroshima, Hiroshima 739-8526, Japan}
\altaffiltext{12}{School of Physics and Technology, Wuhan University, Wuhan, China 430072}

\begin{abstract}

To investigate the dynamics of the galaxy cluster A2142, we compile an extended catalog of 2239 spectroscopic redshifts of sources, including newly measured 237 redshifts, within 30 arcmin from the cluster center. With the $\sigma$-plateau algorithm from the caustic method,
we identify 868 members and a number of substructures in the galaxy distribution both in the outskirts, out to $\sim$3.5 Mpc from the cluster center, and in the central region. In the outskirts,
one substructure overlaps a falling clump of gas previously identified in the X-ray band. These substructures suggests the presence of multiple minor mergers, which are responsible for the complex dynamics of A2142, and the absence of recent or ongoing major mergers. We show that the distribution of the galaxies in the cluster core and in several substructures  are consistent with the mass distribution inferred from the weak lensing signal.
 Moreover, we use spatially-resolved X-ray spectroscopy to measure the redshift of different regions of the intracluster medium within $\sim$3 arcmin from the cluster center. We find a ring of gas near the two X-ray cold fronts identified in previous analyses and measure a velocity of this ring of $810 \pm 330 {\rm km s^{-1}}$ larger than the cluster mean velocity. Our analysis suggests the presence of another ring surrounding the core, whose velocity is $660 \pm 300 {\rm km s^{-1}}$ larger than the cluster velocity. These X-ray features are not associated to any optical substructures, and support the core-sloshing scenario suggested in previous work. \footnote{We dedicate this paper to the late Bepi Tormen, our beloved friend and colleague whose enthusiastic and intense work on gravitational dynamics largely contributed to our current understanding of the formation and evolution of galaxy clusters.}

\end{abstract}

\keywords{galaxies: clusters: general, galaxies: clusters: individual (Abell 2142),
galaxies: clusters: intracluster medium }

\section{Introduction}

Galaxy clusters link the evolution of the large-scale structure to the astrophysical processes on smaller scales, and the study of their assembling is thus
crucial to understand the hierarchical evolution of the universe. The commonly accepted scenario is that clusters form and evolve via accretion and merging of smaller halos. This scenario is suggested
by many dynamical features observed in clusters: substructures in the galaxy distribution  \citep{1982Geller, 2013Wen,guennou2014,girardi2015,zarattini2016}; apparent global rotation of clusters \citep{hwang2007,manolopoulou2017}; clumpy distributions \citep{gutierrez2005,Parekh2015,yu2016,parekh2017} and bow shocks \citep{markevitch2002,markevitch2005} in the intracluster medium (ICM) observed in X-rays; the elongated or peculiar distributions of radio emission \citep{feretti2012,govoni2012,riseley2017,rajpurohit2018}; the substructure distribution of the dark matter, inferred from gravitational lensing observations \citep{okabe2008,okabe2014,grillo2015,caminha2017}. In addition, ``cold fronts"  are frequently observed in X-ray images of clusters \citep{markevitch2000,sanders2005,sanders2016,ichinohe2017}, including some regular and relaxed clusters \citep{mazzotta2001,clarke2004}. Cold fronts are X-ray surface brightness edges with approximately continuous pressure profile across the density discontinuity, at odds with the large pressure jump of shock fronts.

The massive cluster A2142 is one of the most representative clusters with cold fronts. Its {\sl Chandra} image exhibits an elongated X-ray morphology, and two prominent cold fronts in the opposite directions along the longest axis \citep{markevitch2000}. The scenario of A2142, suggested by these observational features, was at first envisioned as a pericentric merging of two subclusters, with the two cold fronts delineating the subcluster cores that have survived the merging. This hypothesis was denoted the ``remnant core" scenario \citep{markevitch2000}, and is appropriate for merging clusters with prominent signatures of recent or ongoing mergers. However, A2142 shows an almost regular morphology and appears relaxed at large radii, unlike 1E 0657-56 \citep{markevitch2002}, A520 \citep{govoni2001,markevitch2005,deshev2017}, and other clusters with cold fronts, which clearly appear unrelaxed. Therefore,
\citet{tittley2005,markevitch2007,owers2011} proposed an alternative model, where the observed cold fronts derive from  a sloshing cool core \citep{markevitch2001,churazov2003}.

More recently, another cold front in A2142, at about 1 Mpc from the center to the southeast, was discovered by \citet{rossetti2013} with XMM-{\sl Newton} observations, showing that the sloshing in A2142 is not confined to the core, but extends to much larger scales. In addition, both XMM-{\sl Newton} \citep{2014Eckert} and {\sl Chandra} \citep{eckert2017} detected a falling clump of hot gas in the outskirts, suggesting that the merging process is still ongoing. Two giant radio halos involved with the sloshing of the cluster core were also revealed by LOFAR and VLA observations \citep{venturi2017}.

Additional information on the complex dynamics of A2142 derives from the optical band.
The presence of substructures in the galaxy distribution was first pointed out by \citet{1995Oegerle} with 103 spectroscopically confirmed galaxies. Based on the spectroscopic redshifts of 956 member galaxies, \citet{owers2011} concluded that some earlier minor mergers can have induced the sloshing of the core in A2142. More recently, \citet{einasto2017} suggested that A2142 formed through past and present mergers of smaller groups, determining the complex radio and X-ray structure observed in this cluster.

All these increasingly rich observational data in different bands provide relevant details of the dynamics of A2142 that, when combined, can further clarify the scenario of the formation of A2142. Here, we provide a step forward in this direction, by combining the information provided by the optical and X-ray spectroscopy and by the dark matter distribution reconstructed from weak-lensing data. The three methods are complementary, and their combination helps to better understand the assembly history of the cluster and to understand the possible systematic errors of
each method (see, e.g., \citealt{hwang2014} and \citealt{yu2016}).

The hierarchical tree method based on optical spectroscopy for the investigation of substructures in galaxy clusters was introduced by \citet{1996Serna}. \citet{1999Diaferio} and \citet{Serra2011} improved this approach by developing the $\sigma$-plateau algorithm for the automatic identification of a threshold to trim the hierarchical tree and identify the cluster substructures. The $\sigma$-plateau algorithm is part of the caustic method \citep{1997Diaferio, 1999Diaferio} that estimates the mass profile of galaxy clusters out to regions well beyond the virial radius \citep{Serra2011} and identifies the galaxies members of the cluster \citep{2013Serra}.  \citet{Yu2015} investigate in detail the performance of the $\sigma$-plateau algorithm
to identify the substructures in the distribution of the cluster galaxies.
Substructure properties, including their redshift, velocity dispersion, and morphology, can provide relevant information on the dynamics of the cluster.

In addition to the optical information, fitting the position of the iron $K_{\alpha}$ line in the X-ray spectrum coming from a defined region of the ICM provides an accurate measure of its redshift \citep{dupke2001,dupke20012,dupke2006,aliu2015}.
By measuring the redshift of different ICM regions,
one can infer a map of the radial velocities of the ICM,
which, again, provides constraints on the modelling of the ICM dynamics:
with this technique, \citet{aliu2016} identified the presence of significant bulk motions
in the ICM of A2142 at a 3$\sigma$ confidence level.

The optical and spatially-resolved X-ray spectroscopy were successfully combined for the first time by \citet{yu2016} for the cluster A85, to unveil the origin of its complex accretion process (see also \citealt{song2017} for A2199).
Here, we apply this new strategy to investigate the dynamics of A2142.

The paper is organized as follows. In Section \ref{sec2}, we describe
our new and extended optical spectroscopic catalog; In Section \ref{sec3}, we identify the substructures
in the galaxy distribution; In Section \ref{sec4}, we describe the
spatially-resolved ICM redshift measurements. In Section \ref{sec5}, we present the method and results of the weak-lensing analysis, and in Section \ref{sec6}, we infer the dynamical state of A2142 by combining
the information coming from our analyses of all the data.
We conclude in Section \ref{sec7}.
Throughout this paper, we adopt the 7 years {\sl WMAP} cosmology, with
$\rm \Omega_{m}$= 0.272, $\rm \Omega_{\Lambda}$ = 0.728, and $H_{0}$
= 70.4 km\ $\rm s^{-1}$\ $\rm Mpc^{-1}$
\citep{komatsu2011}.  All the errors we mention are 1$\sigma$
confidence level.

\section{Optical spectroscopic sample }
\label{sec2}

We compile the currently largest catalog of 2239 spectroscopic redshifts in the field of view of A2142.
Our catalog covers an area of $1^{\circ} \times 1^{\circ} $ around the cluster center, corresponding
to an area  $7\times 7$ Mpc$^2$ at the cluster redshift $z=0.09$;
the redshifts in our catalog are in the range $[0.01,0.6]$.
Our catalog includes 1270 redshifts from the catalog of \citet{owers2011} which are not in the SDSS DR13 \citep{SDSSdr13}; Owers et al.'s full catalog lists 1635 redshifts in the range $z=[0.0088,3.8]$.
In our sample, we include 731 additional redshifts from SDSS catalog and one redshift from \citet{1995Oegerle}. In June 2014, to secure more redshifts, we made additional spectroscopic observations with the 300 fiber Hectospec multi-object spectrograph on the MMT 6.5m telescope \citep{fabricant2005}. To obtain a high, uniform spectroscopic completeness in the cluster region, we weighted the targets according to their r-band apparent magnitudes independently of colors. We used the 270 line $\rm mm^{-1}$ grating for Hectospec observations, which gives a dispersion of 1.2~$\AA~{\rm pixel}^{-1}$ and a resolution of $\sim6~\AA$. The resulting spectra cover the wavelength range 3650--9150 $\AA$. We observed one field with $\sim$250 target fibers for 3$\times$20-minute exposures. The spectra were reduced with the \citet{mink2007} pipeline, and were cross-correlated with template spectra to determine the redshifts using RVSAO \citep{kurtz1998}. We visually inspected all the spectra, and assigned a quality flag to the spectral fits with ``Q" for high-quality redshifts, ``?" for marginal cases, and ``X" for poor fits. We then use only the spectra with reliable redshift measurements (i.e., ``Q"). In total, we obtained 237 additional redshifts in the field.

\citet{owers2011} and \citet{2016Geller,geller2014} show that there is basically no systematic bias, between MMT and SDSS sources; therefore, it is appropriate to merge the two data sets in the same catalog. As mentioned above, the catalog of \citet{owers2011} include redshifts that also appear in the SDSS catalog: we always choose the SDSS measures in this case, because their errors are smaller.

\begin{figure}
\centering
\includegraphics[width=0.48\textwidth]{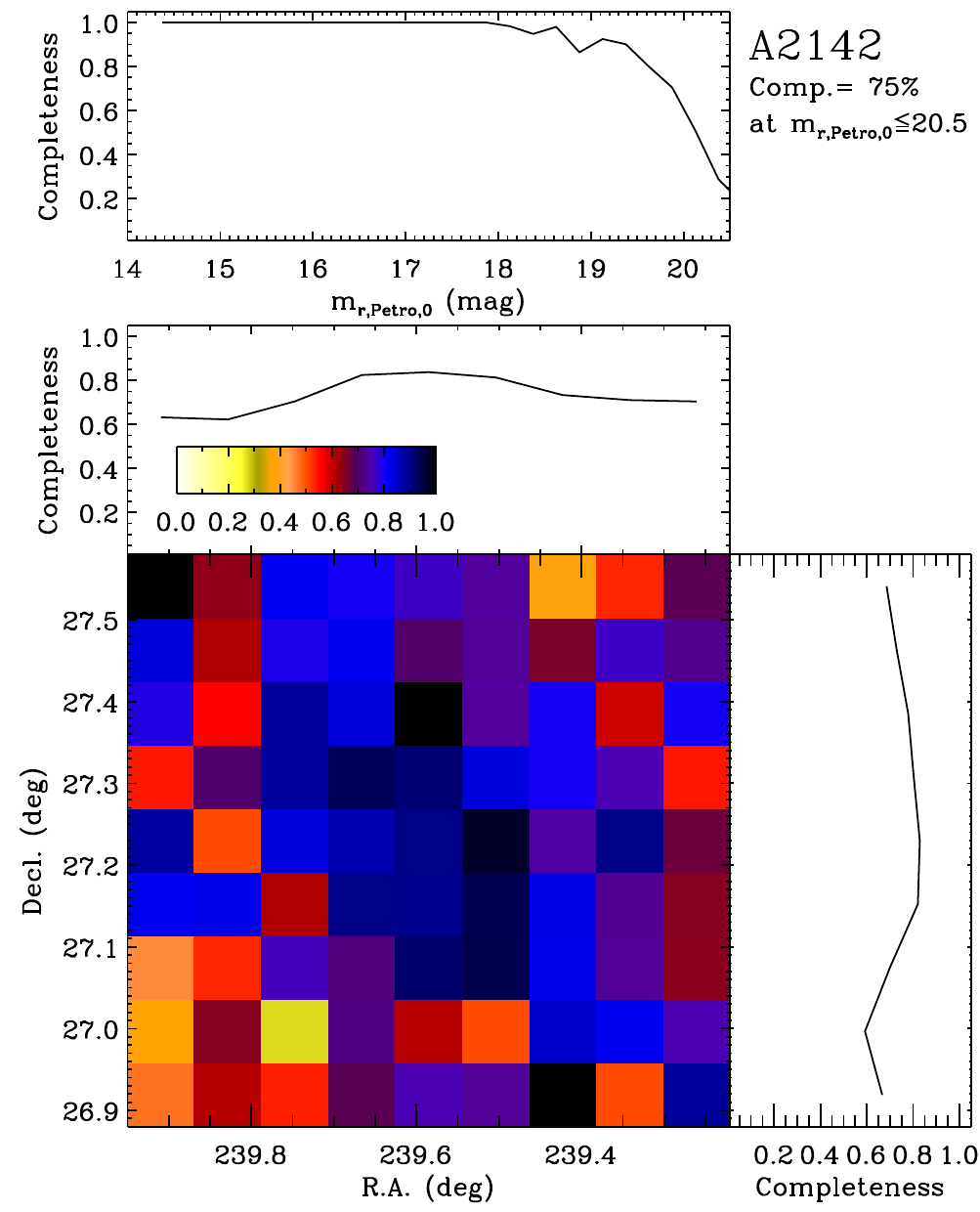}
\caption{Spectroscopic completeness in the field of A2142.}
\label{fig:comp}
\end{figure}

In Figure \ref{fig:comp}, we show the spectroscopic completeness for $m_{\rm r,Petro,0} \le 20.5$. The two-dimensional completeness map is in 9$\times$9 pixels for the $42' \times 42'$ field of view. The overall completeness throughout this field is 75\%, with a small spatial variation.

A sample of the redshift catalog is given in Table \ref{tab-gal}. For each galaxy, the table contains the SDSS ObjID, right ascension R.A., declination  Dec., r-band Petrosian magnitude with  Galactic extinction correction from SDSS, the redshift $z$, the uncertainty in $z$, and the spectrum and redshift source. The full version of the table is available in the online journal.

\begin{figure}
\centering
\includegraphics[width=0.48\textwidth]{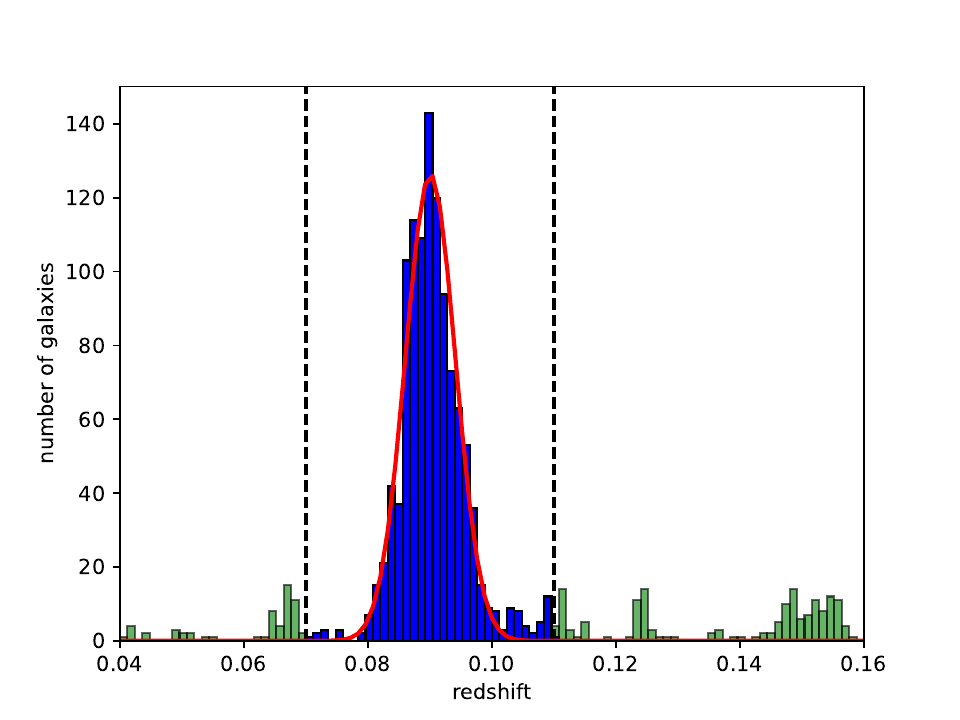}
\caption{The redshift histogram of A2142. The green bars show the distribution
of the field galaxies. The blue bars show the galaxies whose redshifts
range from 0.07 to 0.11 (vertical dashed lines). The red solid line is the
Gaussian fit result after the 3$\sigma$ clipping \citep{yahil1977}.}
\label{fig:vhis}
\end{figure}

Our catalog contains 604  redshifts in addition to the \citet{owers2011} catalog.
Figure \ref{fig:vhis} shows the distribution of the redshifts in our catalog. There are 1117 redshifts in the range $z=[0.07,0.11]$, out of which 63 were not in the \citet{owers2011} catalog.  The $3\sigma$ clipping procedure \citep{yahil1977} removes 50 galaxies and thus leaves 1067 galaxies as possible cluster members. The mean redshift and the redshift dispersion of these 1067 galaxies are $0.0901 \pm 0.0001$ and 0.0040, respectively.

\begin{deluxetable*}{rccccccc}
\tabletypesize{\footnotesize}
\tablewidth{17cm} 
\tablecaption{Redshift catalog in the fields of A2142
\label{tab-gal}}
\tablehead{
ID & SDSS ObjID & R.A.$_{2000}$ & Dec.$_{2000}$ & $m_{\rm r,Petro,0}$ & $z$ & $u_{z}$ & $z$ source\tablenotemark{b}  \\
   & (DR13)     & (deg)         & (deg)         & (mag)               &     &         &
}
\startdata
  26 &  1237662305125859528 & 239.041210 & 27.002581 & 16.950 &  0.06760 & 0.00003 &  SDSS \\
  27 &  1237662339475571013 & 239.041485 & 27.233463 & 17.867 &  0.18976 & 0.00007 &  MMT  \\
  28 &  1237662339475571047 & 239.042569 & 27.177706 & 16.879 &  0.09650 & 0.00001 &  SDSS \\
  29 &  1237662339475571048 & 239.045342 & 27.174775 & 20.041 &  0.09611 & 0.00019 &  Owers \\
  30 &  1237662305662664876 & 239.046669 & 27.442754 & 18.429 &  0.17523 & 0.00007 &  MMT \\
\multicolumn{8}{c}{......}
\enddata
\tablenotetext{1}{This table is available in its entirety in a machine-readable form in the online journal.
                  A portion is shown here for guidance regarding its form and content.}
\tablenotetext{2}{(1) MMT: This study; (2) Owers: \citet{owers2011}; (3) SDSS: SDSS DR13; (4) Oegerle:\citet{1995Oegerle};}
\tablenotetext{3}{The SDSS ObjIDs starting with "12" and "58" come from SDSS DR13 and DR7, respectively.}
\end{deluxetable*}

\section{Substructures in the galaxy distribution}
\label{sec3}

For a quantitative investigation of the distribution of galaxies, we adopt the $\sigma$-plateau algorithm.
To set up a familiar framework to our results, we also provide an analysis based on the Dressler-Shectman (DS) test,
which is more commonly used in the literature.

\subsection{Results from the $\sigma$-plateau algorithm}
\label{sec3.1}

The $\sigma$ plateau algorithm, implemented within the caustic method \citep{1997Diaferio, 1999Diaferio,Serra2011}, is based on optical spectroscopic data and provides an extremely efficient procedure to identify both the cluster members \citep{2013Serra} and the cluster substructures \citep{Yu2015}. Unlike the DS method, which only suggests the presence of substructures but does not unambiguously identify them, the $\sigma$-plateau algorithm returns a list of the individual substructures and their members.

The method consists in a classical approach to cluster analysis for grouping sets of objects with similar properties. The caustic method groups the galaxies in the field of view in a binary tree according to an estimate of their pairwise binding energy, derived from the projected separation and the line-of-sight velocity difference of each galaxy pair.

\begin{figure*}
\centering
\includegraphics[width=0.98\textwidth]{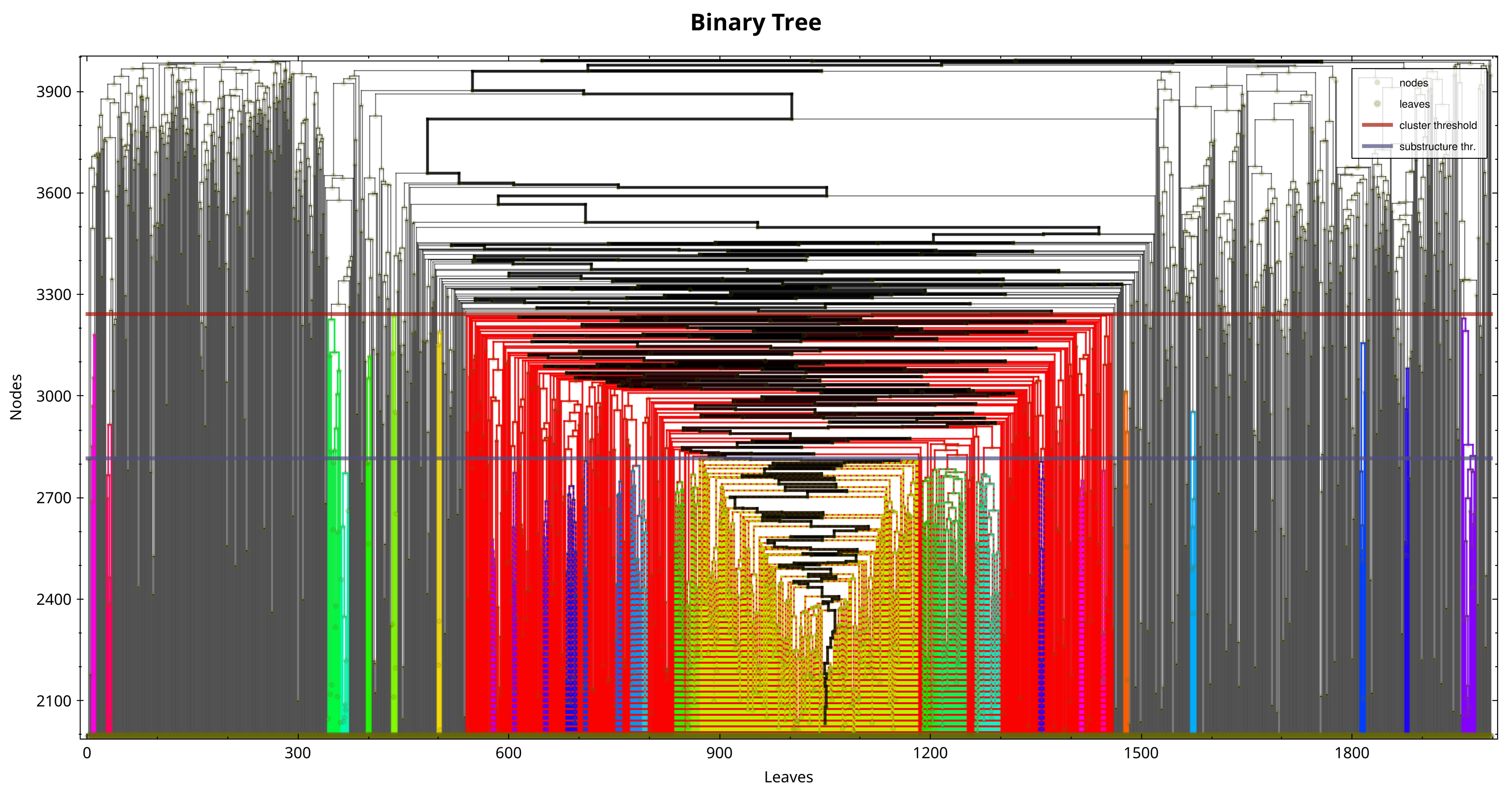}
\caption{The dendrogram tree of 1997 redshifts in the range [0.01, 1.0]. The two solid horizontal lines are the first and second thresholds. The colors indicate different structures. The red part corresponds to sub0 in Table \ref{table:subs}. The heavy black line shows the main branch of the binary tree.}
\label{fig:tree}
\end{figure*}

The main branch of the binary tree is the path determined by the nodes of the tree which contain the largest number of galaxies, or leaves, at each bifurcation. The velocity dispersion $\sigma_i$ of the leaves of each node $i$ decreases, on average, when walking along the main branch from the root to the leaves.

\begin{figure}
\centering
\includegraphics[width=0.48\textwidth]{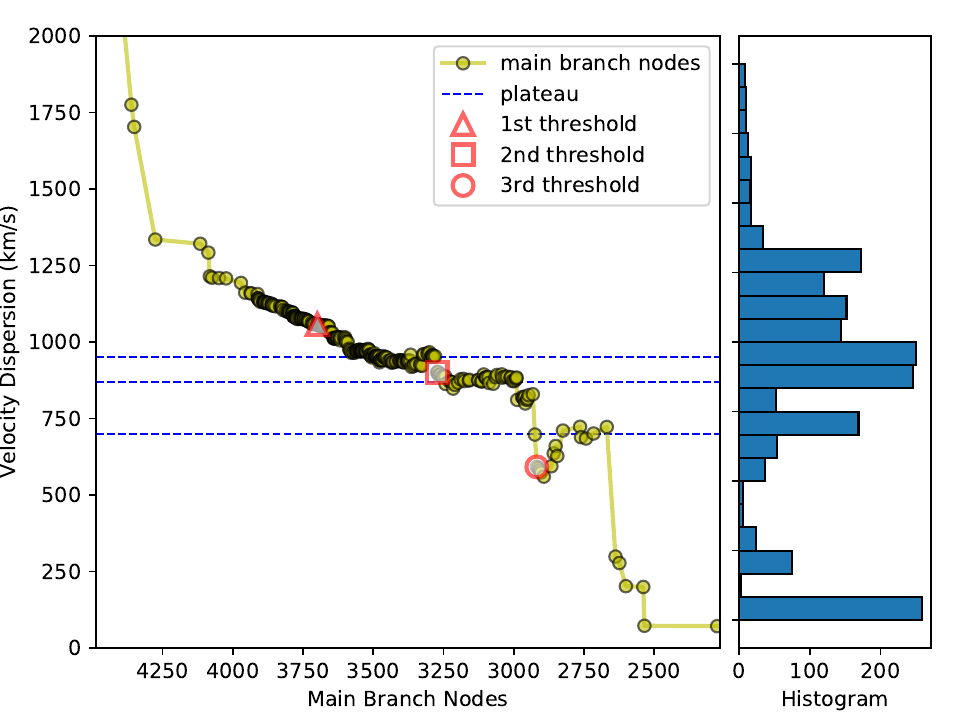}
\caption{The velocity dispersion of the leaves of each node along the main branch
of the binary tree of A2142. The histogram in the right panel shows the
node numbers in different velocity dispersion bins.  The blue dashed
lines indicate the $\sigma$ plateaus.  The red symbols are the selected
thresholds.}
\label{fig:sp}
\end{figure}

When the binary tree is built with the galaxies in a field of view containing a galaxy cluster (see Figure \ref{fig:tree} for our redshift catalog of A2142), the velocity dispersion $\sigma_i$ along the main branch settles onto a $\sigma$ plateau in between two nodes, as shown in Figure \ref{fig:sp}, where we plot the velocity dispersion $\sigma_i$ as a function of the main branch node of A2142. This plateau originates from the quasi-dynamical equilibrium of the cluster: the velocity dispersions $\sigma_i$ of the nodes closer to the root (on the left of Figure \ref{fig:sp}) are larger than the plateau, because these nodes  contain a large fraction of galaxies that are not cluster members; the velocity dispersions of the nodes closer to the leaves (on the right of Figure \ref{fig:sp}) are smaller than the plateau because, at this level, the binary tree splits the cluster into its dynamically distinct substructures.

The two boundary nodes of the plateau thus identify two thresholds that are used to cut the tree at two levels: the threshold closer to the root identifies the cluster members; the threshold closer to the leaves identifies the cluster substructures \citep[see][for further details]{2013Serra,Yu2015}. All the systems with at least 6 galaxies below this second threshold enter our list of substructures. Here, as the minimum number of substructure members, we adopt 6 galaxies, rather than 10 galaxies as chosen by \citet{Yu2015}, which appears to be too severe and can exclude real, albeit poor, substructures.

In the plot of the velocity dispersion along the main branch
of the binary tree of A2142 shown in Figure \ref{fig:sp}
we can see three, rather than one, plateaus: this feature is typical of clusters
with complex dynamics \citep{Yu2015}.
The left node, shown by the open red triangle, of the first plateau at $\sim$950 km$\rm\ s^{-1}$,
is the threshold that identifies the cluster.
The left node, shown by the open red square, of the second plateau at $\sim$870 km$\rm\ s^{-1}$, is
the threshold that identifies the cluster substructures.
The left node of the third plateau at $\sim$700 km$\rm\ s^{-1}$, shown by the open red circle, further splits the
cluster core into additional substructures, as we illustrate below.

\begin{figure}
\centering
\includegraphics[width=0.49\textwidth]{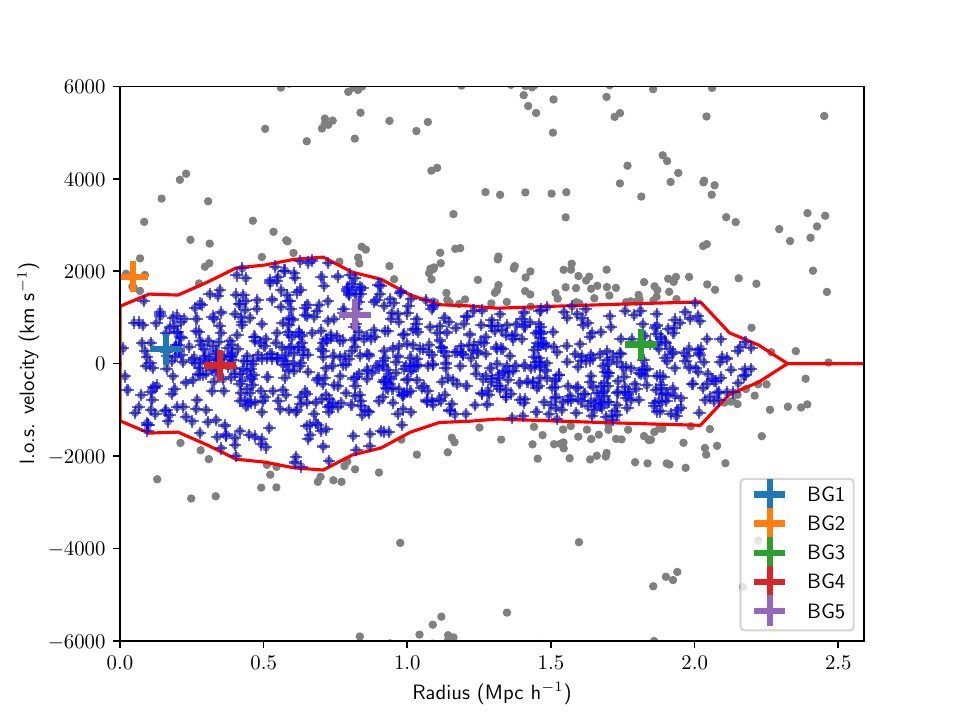}
\caption{The redshift diagram of A2142. The solid lines show the caustic location. The blue crosses show the cluster members identified
by the caustic method. We also show the location of the five brightest galaxies BG1$-$5.}
\label{fig:rediag}
\end{figure}

The caustic method uses the galaxies identified by the first threshold only to locate the
celestial coordinates and the redshift of the cluster center. As discussed by \citet{Serra2011} and \citet{2013Serra}, once the cluster is the dominant system in the field of view, the selection of the cluster
center is not strongly affected by the selection of the first threshold of the binary tree. To identify the cluster members,
the method first locates the caustics in the cluster redshift diagram, the plane
of the line-of-sight velocity of the galaxies and their projected separation from the cluster center (Figure \ref{fig:rediag}). The
caustics measure the galaxy escape velocity from the cluster, corrected by a function of the velocity anisotropy parameter (see \citealt{Serra2011} for details): the galaxies within the caustics are thus the members of the clusters. With this approach, we identify 868 galaxy members of A2142. This number is consistent with, albeit smaller than, the 1067 members identified with the 3$\sigma$ clipping method (see Sect. \ref{sec2}), and the 956 members identified by \citet{owers2011} with the van Hartog and Katgert method \citep{hartog1996}. The difference derives
from the small fraction of interlopers misidentified as members by the caustic method: on average, only 2\% of the caustic
members within the cluster virial radius actually are interlopers and only 8\% within three
times the virial radius are interlopers \citep{2013Serra}.

Table \ref{table:subs} summarizes the basic properties of the substructures,
from sub0 to sub19,
identified by the second threshold at $\sim$870 km$\rm\ s^{-1}$ shown in Figure \ref{fig:sp}. We also list the mean redshifts $z_o$ of the substructures with their uncertainty
${\rm err}_z = v_{\rm disp}/\sqrt{N_{\rm gal}}$, where $v_{\rm disp}$ is the velocity dispersion and ${N_{\rm gal}}$ is the number of members of the substructure \citep{Statistics2014}. Within the substructure sub0, the third threshold at $\sim$700 km$\rm\ s^{-1}$ identifies, at a lower level of the hierarchical clustering, six additional substructures that we list as sub00 to sub05 in Table \ref{table:subs}. To assess whether these substructures correspond to physical systems rather than being chance alignments of unrelated galaxies, below we will combine these results with results from additional probes, including X-ray emission and gravitational lensing measurements.

\begin{figure}
\centering
\includegraphics[width=0.49\textwidth]{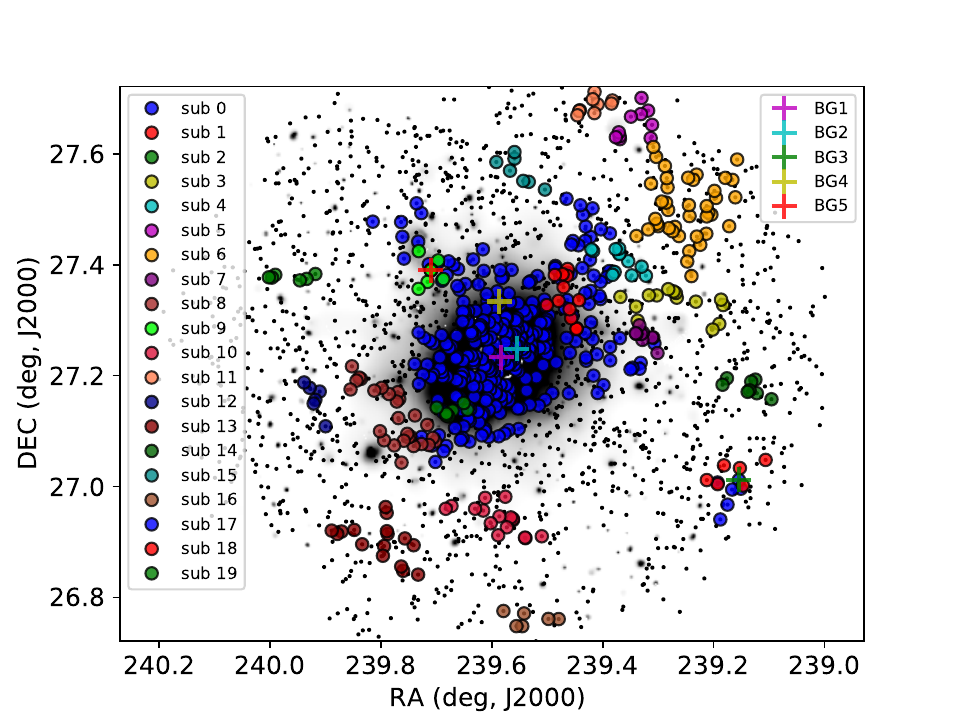}\\
\includegraphics[width=0.49\textwidth]{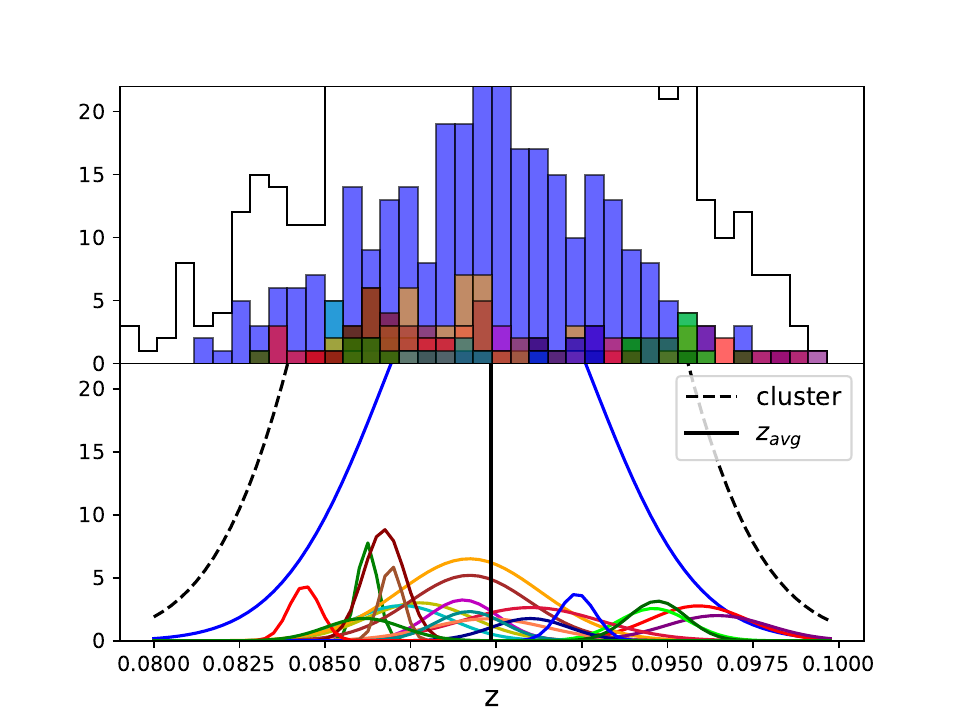}
\caption{The upper panel shows the distribution on the sky of
the A2142 substructures identified with the second threshold. The white contours and grey scale of the background image show the X-ray surface brightness from the combined X-ray observations of XMM \citep{Tchernin2016}}. The middle and bottom panels show the velocity histograms of the substructures and
their best Gaussian fits.  The black vertical line shows the position of
the mean redshift $z_{\rm avg}=0.08982 \pm 0.00010$ as a reference.
\label{thres2}
\end{figure}

\begin{figure}
\centering
\includegraphics[width=0.49\textwidth]{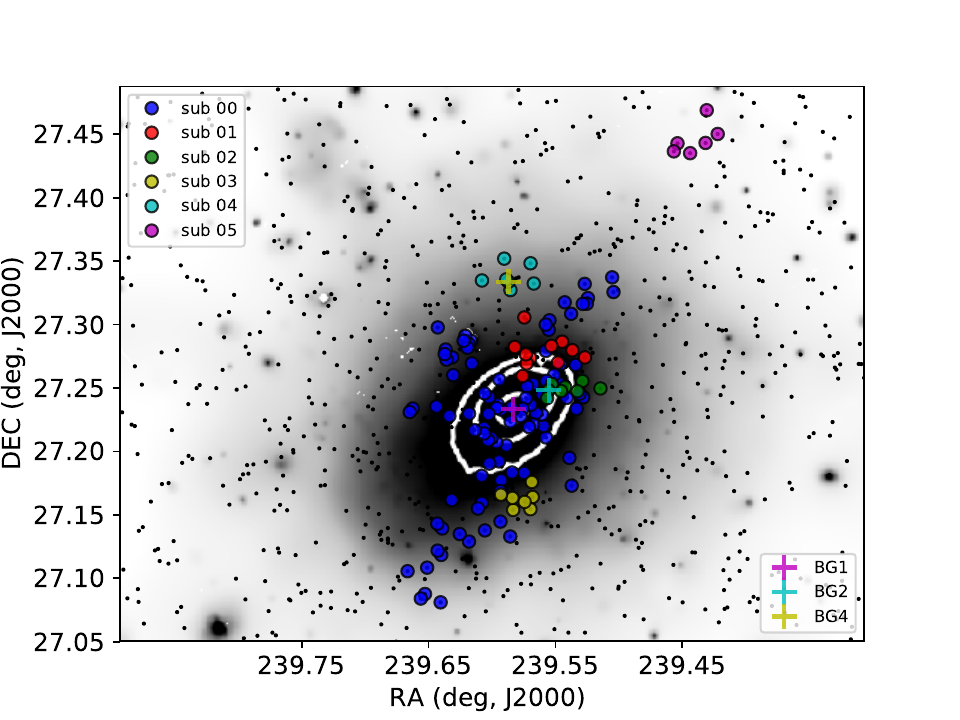}\\
\includegraphics[width=0.49\textwidth]{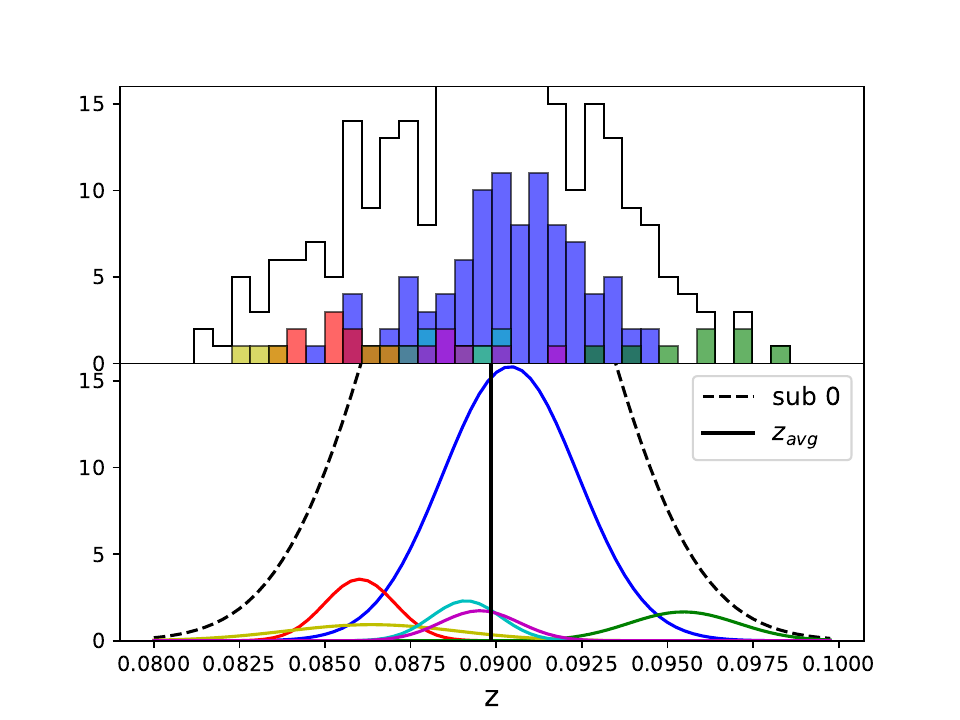}
\caption{Same as  Figure \ref{thres2} for the A2142 substructures identified
with the third threshold.}
\label{thres3}
\end{figure}

 Here, we also consider the relation between these substructures and the top five brightest (r-band Petrosian magnitudes) galaxies in the cluster (labelled as BG1$-$5). They are all cluster members, except BG2. BG1 and BG2 are also well known Brightest Cluster Galaxies.  As shown in Figures \ref{thres2} and \ref{thres3}, the brightest one -- BG1 -- is in sub00, which is the core of the cluster. BG2 is a member of sub02 with a substantial velocity offset: its line-of-sight velocity is 312 km$\rm\ s^{-1}$ larger than the mean velocity of the members of sub02. The spiral galaxy BG3 is a member of sub17, a system in the outskirts of the cluster; unfortunately, this region is spectroscopically severely undersampled and we
are unable to draw any solid conclusion. Finally, the elliptical galaxy BG4 and the interacting galaxy BG5 are members of sub04 and sub9
respectively. We will discuss the properties of sub9 and BG5 in section \ref{sec6}.

\begin{deluxetable}{cccc}
\centering
\tablewidth{0.45\textwidth}
 \tablecaption{Physical properties of the substructures}
 \tablehead{ \colhead{ GroupID } & \colhead{$N_{\rm gal}$}   & \colhead{$z_{o}$}  & \colhead{$v_{\rm disp}\;$(km~s$^{-1})$} }
\startdata  cluster & 868 & 0.08982 $\pm$ 0.00010 & 902 $\pm$ 22\\
\hline
sub0& 311 & 0.08977 $\pm$ 0.00017 & 901 $\pm$ 36\\
sub1& 12 & 0.09590 $\pm$ 0.00042 & 431 $\pm$ 91\\
sub2& 7 & 0.08623 $\pm$ 0.00011 & 89 $\pm$ 25\\
sub3& 14 & 0.08776 $\pm$ 0.00041 & 462 $\pm$ 90\\
sub4& 12 & 0.08725 $\pm$ 0.00041 & 428 $\pm$ 91\\
sub5& 10 & 0.08903 $\pm$ 0.00032 & 307 $\pm$ 72\\
sub6& 40 & 0.08922 $\pm$ 0.00032 & 612 $\pm$ 69\\
sub7& 9 & 0.09645 $\pm$ 0.00050 & 447 $\pm$ 111\\
sub8& 27 & 0.08923 $\pm$ 0.00033 & 517 $\pm$ 71\\
sub9& 8 & 0.09459 $\pm$ 0.00037 & 310 $\pm$ 83\\
sub10& 16 & 0.09101 $\pm$ 0.00050 & 604 $\pm$ 110\\
sub11& 9 & 0.08976 $\pm$ 0.00057 & 513 $\pm$ 128\\
sub12& 6 & 0.09099 $\pm$ 0.00046 & 337 $\pm$ 106\\
sub13& 16 & 0.08672 $\pm$ 0.00015 & 180 $\pm$ 32\\
sub14& 8 & 0.09470 $\pm$ 0.00030 & 253 $\pm$ 67\\
sub15& 7 & 0.08921 $\pm$ 0.00038 & 300 $\pm$ 86\\
sub16& 6 & 0.08693 $\pm$ 0.00014 & 100 $\pm$ 31\\
sub17& 6 & 0.09235 $\pm$ 0.00022 & 160 $\pm$ 50\\
sub18& 6 & 0.08440 $\pm$ 0.00019 & 136 $\pm$ 43\\
sub19& 6 & 0.08614 $\pm$ 0.00046 & 335 $\pm$ 106\\
\hline
sub00& 94 & 0.09041 $\pm$ 0.00020 & 591 $\pm$ 43\\
sub01& 11 & 0.08601 $\pm$ 0.00031 & 308 $\pm$ 68\\
sub02& 8 & 0.09547 $\pm$ 0.00056 & 478 $\pm$ 127\\
sub03& 7 & 0.08636 $\pm$ 0.00094 & 748 $\pm$ 216\\
sub04& 7 & 0.08911 $\pm$ 0.00038 & 302 $\pm$ 87\\
sub05& 6 & 0.08952 $\pm$ 0.00047 & 344 $\pm$ 108
\enddata
\label{table:subs}
\end{deluxetable}

\subsection{Results from the Dressler-Schectman test}
\label{sec3.2}
The Dressler-Shectman (DS) test is largely used for the investigation of cluster substructures when
optical spectroscopic data are available \citep{1988Dressler,1996Pinkney}. The test requires
the identification of the $N$ cluster members, which is usually obtained by removing the possible
interlopers with the 3$\sigma$ clipping procedure \citep{yahil1977}. Around each cluster member $i$, we identify the $N_{\rm local}=\sqrt{N}$ closest neighbors whose mean velocity $v_{\rm local}^i$ and velocity dispersion $\sigma_{\rm local}^i$ is compared with the mean velocity $v$ and velocity dispersion  $\sigma$ of the $N$ members of the entire cluster. We thus define a local kinematic deviation for each
cluster member
\begin{equation}
\delta_i^2 = \frac{N_{\rm local}+1}{\sigma^2} [(v_{\rm local}^i-v)^2+(\sigma_{\rm local}^i-\sigma)^2]\; .
\end{equation}
The cumulative deviation $\Delta = \Sigma_i \delta_i $ is used as the test statistic to quantify the statistical significance of the presence of substructures.
For a cluster with a Gaussian distribution of the member velocities, $\Delta$ is close to $N$. If the velocity distribution deviates from a Gaussian, $\Delta$ could vary significantly from $N$, either with or without substructures. Therefore, the statistical significance of the presence of substructures can be quantified by the ratio
$p \equiv N(\Delta_{\rm simu}>\Delta_{\rm obs})/N_{\rm simu}$, where $\Delta_{\rm simu}$ is the value of $\Delta$ estimated in $N_{\rm simu}$ Monte Carlo simulations where the  velocities of the galaxies are randomly shuffled while the galaxy celestial coordinates are kept fixed, and $N(\Delta_{\rm simu}>\Delta_{\rm obs})$ is the number of simulations where $\Delta_{\rm simu}>\Delta_{\rm obs}$, where $\Delta_{\rm obs}$ is the value obtained from the original data set. A small $p$ thus suggests a significant presence of substructures.

\begin{figure}
\centering
\includegraphics[width=0.5\textwidth, trim=25 0 0 0, clip]{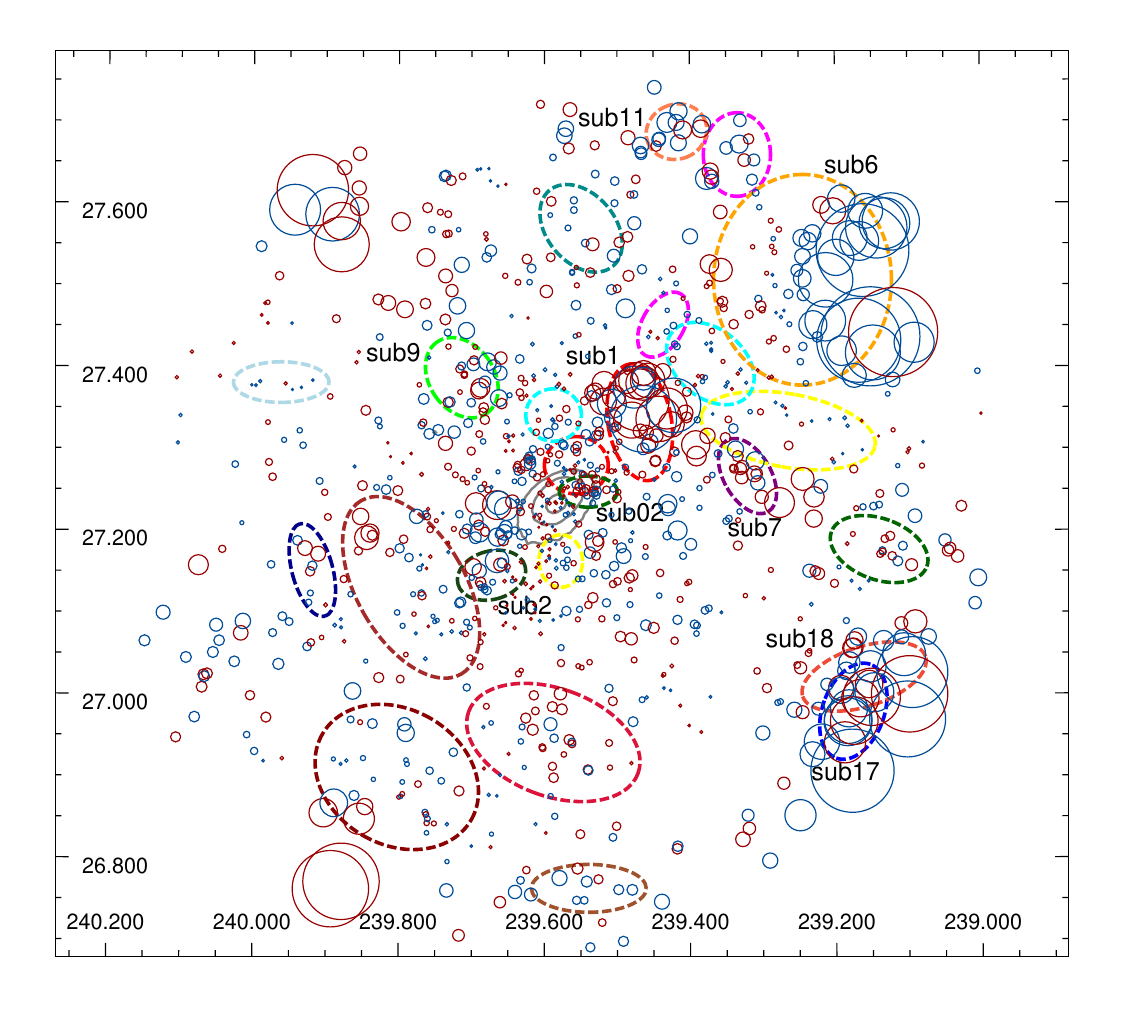}
\caption{Results of the DS analysis based on the 1067 member galaxies identified by the 3$\sigma$ clipping procedure on galaxy velocities. Blue and red circles represent galaxies with velocities smaller and larger than the cluster mean velocity respectively. The radius of each circle is proportional to $e^{\delta_i}$. The ellipses mark the substructures identified with the $\sigma$-plateau algorithm. The color of these substructures are the same as in Figure \ref{thres2} and \ref{thres3}. We only label with their names the substructure ellipses that overlap a clump of DS circles. Grey contour levels in the center show the distribution of the X-ray emission.  }
\label{fig:ds}
\end{figure}

We apply the DS test to our A2142 catalog. With $N=1067$ and $N_{\rm local}=33$, we obtain
$\Delta_{\rm obs}/N = 1.46$. We run $N_{\rm simu}=10000$ simulations and obtain $p = 0.001$, which strongly indicates the existence of substructures. Figure \ref{fig:ds} shows the $\delta_i$ of each cluster member on the plane of the sky. The radius of each circle is proportional to $e^{\delta_i}$. Blue and red circles represent galaxies with smaller and larger peculiar velocities with respect to the cluster mean velocity respectively. The clustering of circles with similar radii therefore suggests the presence of substructures. We stress that the large circles close to the edge of the field are unreliable, because the cluster outskirts are spectroscopically severely undersampled.

We now compare the results of the $\sigma$-plateau algorithm with the DS analysis. The areas including all the members of each substructure sub1--19, and sub01--05 are delimited by ellipses on Figure \ref{fig:ds}. We label sub1, 2, 6, 7, 9, 11, 17, 18, and sub02: they overlap clumps of circles derived from the DS analysis. Sub5, the purple structure in the NE, also overlaps a DS structure, but it is less secure because it is on the border of the field of view where the spectroscopic catalog is largely incomplete.

We note that the DS method looks for galaxy neighbors on the plane of the sky; therefore, clumps of galaxies that might be associated to real substructures might also contain cluster members that have velocities close to the cluster mean velocity. This event can commonly result in clumps of galaxies that are not plotted with the same color. A typical example is the DS clump at (RA=$239.460$, $\delta$=$27.33$) that overlaps sub1.

\section{Spatially-resolved ICM Redshift Measurements}
\label{sec4}

In this section, we investigate the dynamics of the ICM in the center of the cluster with spatially resolved X-ray redshift measurements.  There are numerous observations of A2142 from XMM and {\sl Chandra}. Since spatial resolution is crucial in this work, we focus on the {\sl Chandra} data. In the {\sl Chandra} data archive, we find seven observations available, as listed in Table \ref{xlist}. For X-ray redshift measurements, we only select the three most recent observations taken with ACIS-S to minimize possible systematic errors deriving from different calibrations. The total exposure time is 155.1 ks after data processing.  After examining the stacked image, we select a circle of $\sim$3 arcmin for our spatially resolved spectral analysis. The total number of net counts in the 0.5--10 keV energy band
within this region is $\sim 10^6$.

\begin{deluxetable}{cccc}
\tablewidth{0.48\textwidth}
\tablecaption{List of XMM and {\sl Chandra} observations of A2142. }
\tablehead{ \colhead{CCD} & \colhead{ObsID} & \colhead{Exptime (ks)}   & \colhead{Date} }
\startdata
EPIC & 0111870101 & 35.5 & 2002-07-20 \\
EPIC & 0111870401 & 13.7 & 2002-09-08 \\
EPIC & 0674560201 & 59.4 & 2011-07-13 \\
ACIS-S & 1196 & 11.4 & 1999-09-04 \\
ACIS-S & 1228 & 12.1 & 1999-09-04 \\
ACIS-I & 5005 & 44.5 & 2005-04-13 \\
ACIS-I & 7692 & 5.0 & 2007-05-07 \\
\textbf{ACIS-S} & \textbf{15186} & \textbf{89.9} & \textbf{2014-01-19} \\
\textbf{ACIS-S} & \textbf{16564} & \textbf{44.5} & \textbf{2014-01-22} \\
\textbf{ACIS-S} & \textbf{16565} & \textbf{20.8} & \textbf{2014-01-24}
\enddata
\tablenote{The three most recent observations we used for the X-ray redshift measurements are highlighted.}
\label{xlist}
\end{deluxetable}

\begin{figure}
\centering
\includegraphics[width=0.5\textwidth, trim=10 0 0 0, clip]{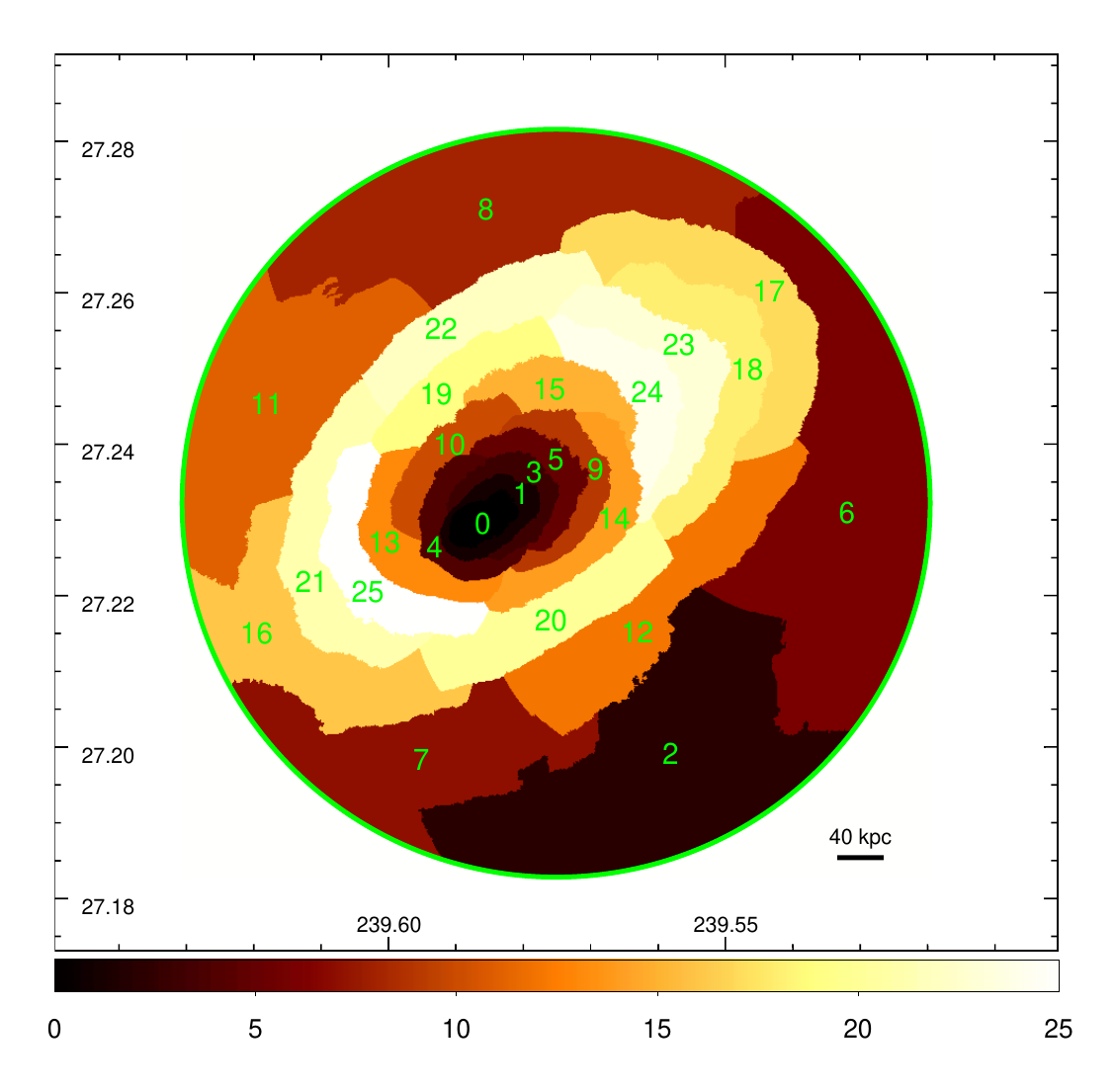}
\caption{The region map produced on the basis of the surface brightness contours. The color bar denotes the serial number of the regions and has no physical meaning.}
\label{regmap}
\vfill
\end{figure}

Similarly to \citet{aliu2016}, we apply the Contour Binning technique \citep{Sanders2006}
to separate the cluster field of view into independent regions based on the surface brightness contours.
In order to acquire a more reliable redshift measurement in each region, we adopt a threshold of signal-to-noise ratio S/N larger than that in \citet{aliu2016}.
We separate the circle of 3 arcmin around the cluster center into 26 regions with S/N larger than 200. Figure \ref{regmap} shows the map of these regions.

The spectra are fitted with {\sl Xspec} v12.9.1 \citep{1996Arnaud}
in the 0.5--10 keV band. Because the chip S3 is entirely covered by the cluster emission, we extract the background from the chip S1. We also check the background modeled from the ``blank sky" dataset, and find that the two results are consistent. To model the X-ray emission from a projected
region, we use the double-{\tt apec} thermal plasma emission model \citep{smith2001}.
The double-temperature thermal model is helpful to reduce the possible
bias in the measurement of the iron line centroid due to the presence of
unnoticed thermal components along the line of sight \citep{aliu2015}. Galactic absorption
is described by the model {\tt tbabs} \citep{2000Wilms}.  The ICM
temperature, the metallicity, the redshift, and the
normalization are all set unconstrained at the same time.
The redshifts of the two components in the model are always linked.
Considering the large parameter space to explore, for the fitting
we adopt a Monte Carlo Markov Chain (MCMC) method. The
chain is generated by the Goodman-Weare algorithm \citep{goodman2010},
with 10 walkers, $10^{4}$ burn-in steps and the total length of $10^{6}$
steps.  After the fitting, chains are top-hat filtered according to the
following ranges: temperature from 0.1 keV to 25 keV, metallicity from 0.001
to 2, and redshift from 0.05 to 0.15.  The best fit parameters and their
uncertainties are estimated from these filtered chains.
The best fit results of the regions are listed in Table \ref{result}.

\begin{deluxetable}{cccc}
\centering
\tablewidth{0.48\textwidth}
 \tablecaption{X-ray Fitting Results.}
 \tablehead{ \colhead{Region ID}   & \colhead{X-ray redshift}   & \colhead{Region ID}   & \colhead{X-ray redshift}
 }
 \startdata
 00  & $ 0.0834_{-0.0007}^{+0.0007} $  & 13  & $ 0.0906_{-0.0031}^{+0.0051} $  \\
 01  & $ 0.0911_{-0.0017}^{+0.0015} $  & 14  & $ 0.0862_{-0.0049}^{+0.0022} $  \\
 02  & $ 0.0862_{-0.0082}^{+0.0075} $  & 15  & $ 0.0849_{-0.0004}^{+0.0015} $  \\
 03  & $ 0.0911_{-0.0013}^{+0.0011} $  & 16  & $ 0.0972_{-0.0085}^{+0.0020} $  \\
 04  & $ 0.0872_{-0.0001}^{+0.0025} $  & 17  & $ 0.0892_{-0.0007}^{+0.0032} $  \\
 05  & $ 0.0866_{-0.0025}^{+0.0019} $  & 18  & $ 0.0954_{-0.0023}^{+0.0040} $  \\
 06  & $ 0.0877_{-0.0054}^{+0.0055} $  & 19  & $ 0.0845_{-0.0022}^{+0.0008} $  \\
 07  & $ 0.0898_{-0.0194}^{+0.0184} $  & 20  & $ 0.0920_{-0.0034}^{+0.0016} $  \\
 08  & $ 0.0876_{-0.0054}^{+0.0073} $  & 21  & $ 0.0852_{-0.0016}^{+0.0032} $  \\
 09  & $ 0.0879_{-0.0013}^{+0.0015} $  & 22  & $ 0.0913_{-0.0017}^{+0.0016} $  \\
 10  & $ 0.0862_{-0.0013}^{+0.0016} $  & 23  & $ 0.0897_{-0.0033}^{+0.0035} $  \\
 11  & $ 0.0879_{-0.0041}^{+0.0055} $  & 24  & $ 0.0876_{-0.0015}^{+0.0011} $  \\
 12  & $ 0.0873_{-0.0020}^{+0.0011} $  & 25  & $ 0.0931_{-0.0029}^{+0.0030} $  \\

 \enddata
\label{result}
\end{deluxetable}

\begin{figure}
\centering
\includegraphics[width=0.5\textwidth, trim=10 0 0 0, clip]{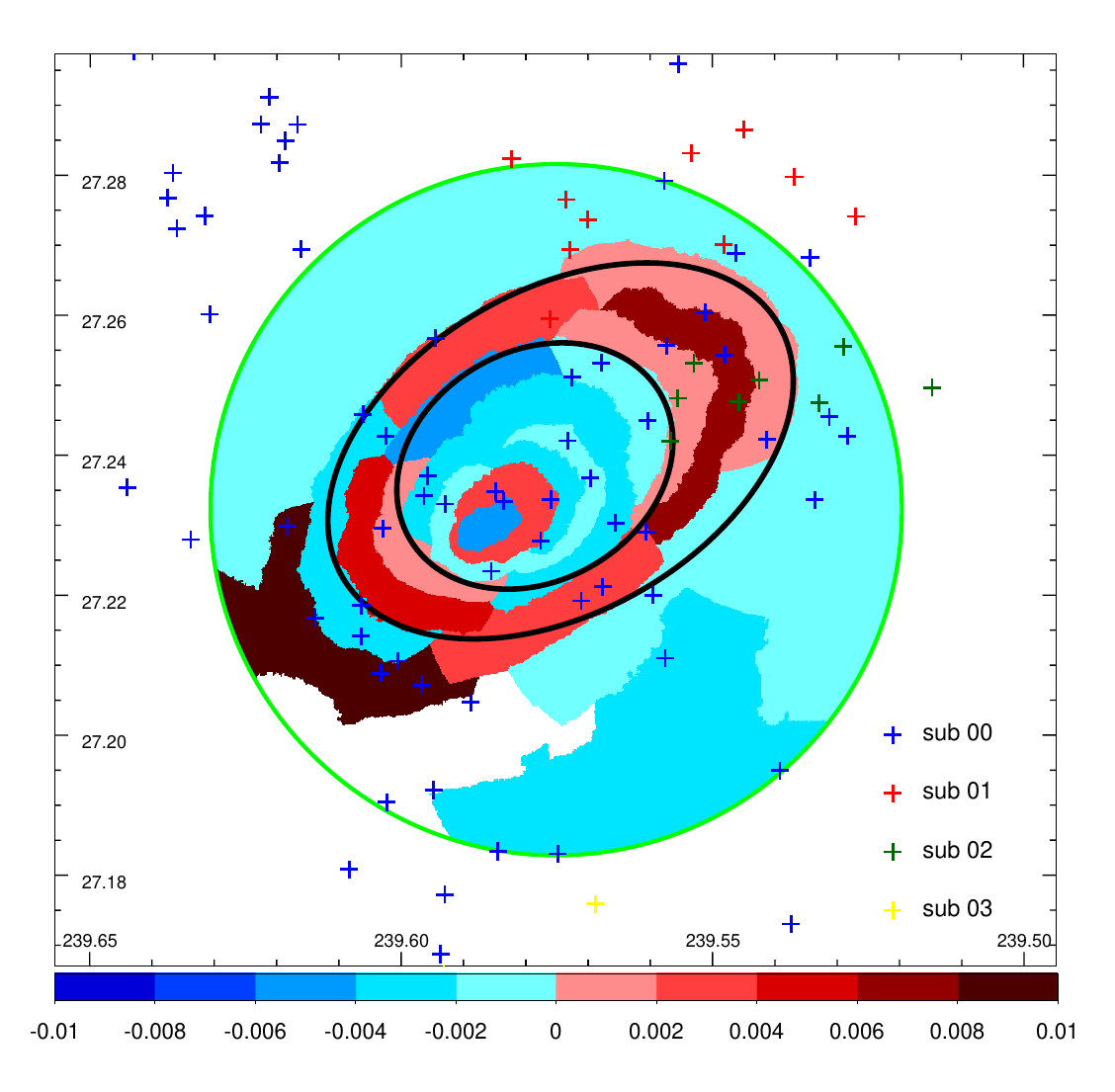}
\caption{Redshift map of A2142. The color bar indicates the redshift difference with respect to the average X-ray redshift $z_{\rm X} = 0.0889\pm0.0009$. The black ellipses roughly mark the elliptical annulus with redshift larger than the surroundings.}
\label{redshift}
\end{figure}

We present our results in the form of a redshift map, shown in Figure \ref{redshift}. Region 7 has a redshift uncertainty larger than 0.01 and appears white. The most prominent feature emerging from the redshift map is that seven regions form an elliptical annulus, marked with the two black ellipses in Figure \ref{redshift}, with a mean redshift larger than its surrounding regions. Specifically, its redshift is $0.0916 \pm 0.0011$. The velocity difference between the annulus and the cluster average is therefore $810 \pm 330$~km s$^{-1}$. The emerging of this high-redshift annulus appears to be consistent with the scenario where the ICM is sloshing due to one or more perturbations. However, the limited spatial resolution of the map implies that we can only derive rough estimates of both the size and the redshift of the annulus.

Additionally, we note that the cluster core is surrounded by another high redshift annulus with redshift $0.0911 \pm 0.0010$. The velocity difference between this annulus and the cluster average is $660 \pm 300 {\rm km s^{-1}}$. A possible interpration of this feature is a small scale sloshing, that generates a ``wave"-like motion of the ICM. Alternatively, this feature could be the signature of the rotation of the cool core, which is an event also suggested by the spiral-like structures observed in other clusters \citep{lagana2010}. Clearly, the projection effects and the large systematic uncertainties in the X-ray redshift measurements require deeper observations with the next generation X-ray bolometers to pin down the appropriate scenario.

\section{Weak lensing data and analysis}
\label{sec5}

\begin{figure}[!htb] 
  \begin{center}
   \includegraphics[scale=0.45, angle=0, trim=0 40 50 0, clip]{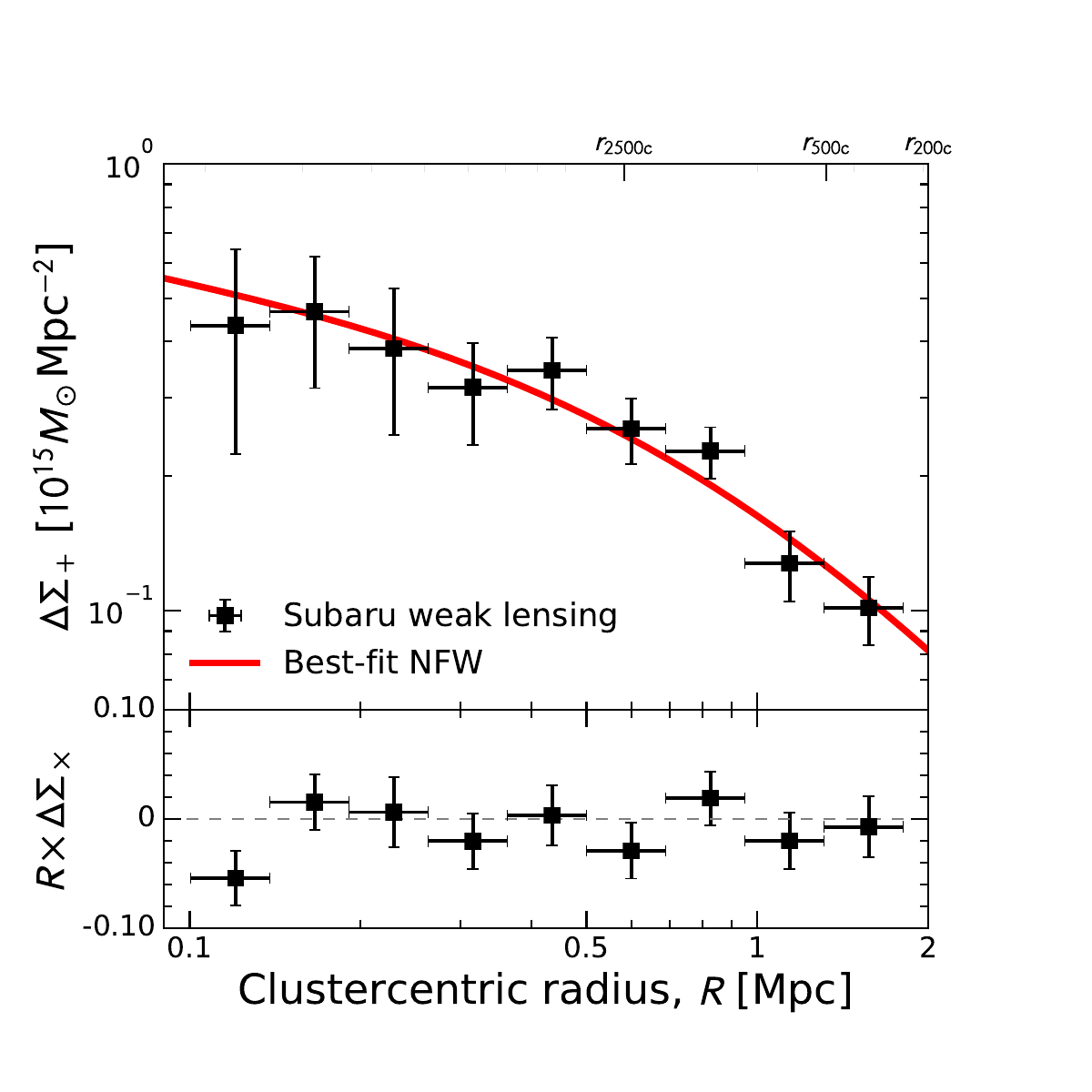}
  \end{center}
\caption{
Tangential reduced shear profile (upper panel, black
 squares) of A2142 derived from our Subaru/Suprime-Cam weak-lensing
 observations, shown
in units of projected mass density. The thick-solid (red) line corresponds to
 the best-fit NFW profile.
 The lower panel shows the $45^\circ$ rotated $\times$ component, which is consistent
with a null signal within $2\sigma$ at all cluster radii. }
 \label{fig:gt}
\end{figure}

A2142 is among the seven nearby merging clusters targeted by the Subaru
weak-lensing analysis of \citet{okabe2008}, who performed a
detailed comparison of the weak-lensing mass distribution with the X-ray
brightness and cluster galaxy distributions.
\citet{umetsu2009} conducted a combined weak-lensing and
Sunyaev-Zel'dovich effect analysis of A2142, along with three other
X-ray luminous clusters targeted by the 7-element AMiBA project
\citep{ho2009} to determine the hot gas fractions in the
clusters in combination with X-ray temperatures.
The \citet{umetsu2009} weak-lensing analysis of A2142 is based on the
same Subaru images as in \citet{okabe2008}, but their improved
method of selecting blue+red background galaxies in $g'R_\mathrm{C}$
color-magnitude space increased the size of the background sample by a
factor of 4 relative to that of \citet{okabe2008}.
With the improved
background selection, \citet{umetsu2009} obtained a virial mass estimate of
$M_\mathrm{vir}=15.2^{+3.1}_{-2.3}\times 10^{14}M_\odot$
and a concentration of $c_\mathrm{vir}=5.5\pm1.1$ (see their
Table 4)
from tangential shear fitting assuming a spherical Navarro--Frenk--White
halo
\citep[][NFW hereafter]{navarro1996}.

Here we revisit the weak-lensing properties of A2142 by performing a
weak-lensing analysis using our most recent shape measurement pipelines
employed by the CLASH collaboration \citep{umetsu2014} and the LoCuSS
collaboration \citep{okabe2016}.
We analyze the Subaru/Suprime-Cam
$g'R_\mathrm{C}$ images reduced by \citet{okabe2008},
and apply the same color-magnitude cuts as presented in
\citet{umetsu2009} to select background galaxies.
We use the Subaru $R_\mathrm{C}$-band images for the shape measurement,
as done in previous work.
Briefly summarizing, the key common feature in our shape measurement
pipelines is that only those galaxies detected with sufficiently
high significance are used to model the isotropic
point-spread-function correction as a function of object size and
magnitude \citep[for details, see][]{umetsu2014,okabe2016}.
For each galaxy, we apply a shear
calibration factor, $g\to g/0.95$, to account for the residual correction estimated
using simulated Subaru/Suprime-Cam images
\citep{umetsu2014,okabe2016}.
All galaxies with usable shape measurements are then matched with those
in the blue+red background samples.
Our conservative selection criteria yield a mean surface number density
of $n_\mathrm{g}\simeq 25$\,galaxies\,arcmin$^{-2}$ for the
weak-lensing-matched background catalog, compared to
$n_\mathrm{g}\simeq 30$\,galaxies\,arcmin$^{-2}$ found by \citet{umetsu2009}.
 We checked that our results from
the two different pipelines \citep{umetsu2014,okabe2016} are
robust and entirely consistent with each other.

In Figure \ref{fig:gt} we show the tangential reduced shear profile in
units of projected mass density,
$\Delta\Sigma_+(R)=\Sigma_\mathrm{c}g_{+}(R)$, with
$\Sigma_\mathrm{c}\simeq 5.5\times 10^{15}M_\odot\,$Mpc$^{-2}$
the critical surface mass density for lensing and $g_+(R)$ the
azimuthally averaged reduced tangential shear as a function of
cluster-centric radius $R$.
We fit the $\Delta\Sigma_+(R)$ profile with a spherical NFW halo using
log-uniform priors for
$M_\mathrm{200c}$ and $c_\mathrm{200c}$ in the range
$0.1<M_\mathrm{200c}/(10^{15}M_\odot\,h^{-1})<10$
and
$0.1<c_\mathrm{200c}<10$, where $h \equiv H_0/(100\,\mathrm{km}~{\rm s}^{-1}\,\mathrm{Mpc}^{-1}) = 0.704$.
The error analysis includes the contribution from cosmic noise due to
the uncorrelated large scale structure projected along the line of sight
\citep{hoekstra2003}, as well as galaxy shape noise and
measurement errors. The mass and concentration parameters are constrained as
$M_\mathrm{200c}=(13.0\pm 2.7)\times 10^{14}M_\odot$ and
$c_\mathrm{200c}=4.1\pm 0.8$, or $M_\mathrm{vir}=(16.1\pm 3.7)\times
10^{14}M_\odot$ and $c_\mathrm{vir}=5.5\pm 1.1$, in good
agreement with the results of \citet{umetsu2009}.  The uncertainties
are slightly larger than those estimated by \citet{umetsu2009}, who
did not account for the cosmic noise contribution.

The mass estimated with the weak lensing analysis is consistent with the mass profile
estimated with the caustic method, based on the amplitude of the caustics in the redshift diagram (Figure \ref{fig:rediag}). The caustic method estimates the radius $r_{200}=2.15$~Mpc and the mass within $r_{200}$, $M_{200}=(11.5\pm 3.7) \times 10^{14}M_\odot$. This agreement confirms the results of \citet{geller2013}, who show that, for a sample of 19 clusters, the caustic and weak lensing masses within $r_{200}$ agree to within $\sim 30$\%, similarly to the early results of \citet{diaferio2005}.

In Figure \ref{wl}, we show the weak-lensing mass map of A2142.
The mass map is smoothed with a Gaussian of $\mathrm{FWHM} =1\arcmin .5$. The mass map exhibits an extended structure elongated along the northwest-southeast direction, consistent with the direction of elongation of the X-ray emission. We compare the mass map with the previous map reconstructed from the old shape catalog \citep{okabe2008}. Since the number density of background galaxies for this analysis is slightly lower than that of the previous analysis, the Gaussian FWHM used for the map is $1\arcmin .5$, larger than that for the old map ($1\arcmin .0$). \citet{okabe2008}  found the main peak at $\sim11\sigma$ and a possible substructure at $\sim3\sigma$ in the northwest region. The noise level was computed with the theoretical estimations from the number density and the variance of ellipticity. Here, we find the main peak at $\sim10\sigma$ and the northwest substructure at $\sim3.3\sigma$. The noise is computed by the bootstrap re-sampling with $3000$ realizations using random ellipticity catalog in order to conservatively evaluate a spatially-dependent noise level caused by sparse galaxy distributions. Although the techniques are revised from the previous study, the overall mass distributions are similar to each other. In this paper, we compare the new map with the distributions of the substructures determined by the dynamical method.

\section{Combined analysis and discussion}
\label{sec6}

\begin{figure}
\centering
\includegraphics[width=0.5\textwidth, trim=25 0 0 0, clip]{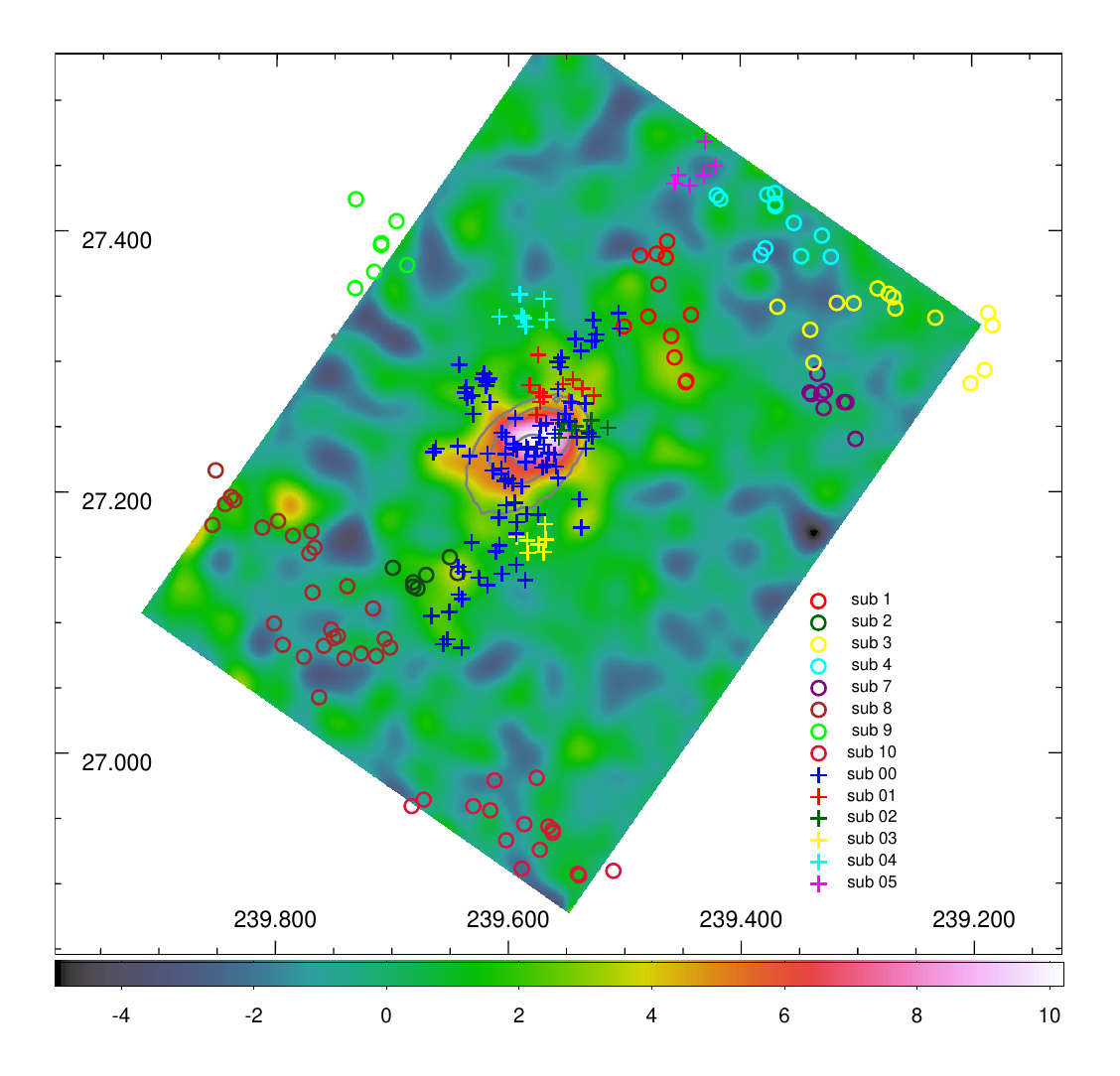}
\caption{The weak-lensing mass map superimposed with the substructures identified with the $\sigma$-plateau algorithm. The color code shows the weak-lensing signal-to-noise ratio, namely
the surface mass density in units of the $1\sigma$ reconstruction error. }
\label{wl}
\end{figure}

\begin{figure}
\centering
\includegraphics[width=0.5\textwidth, trim=60 150 20 200, clip]{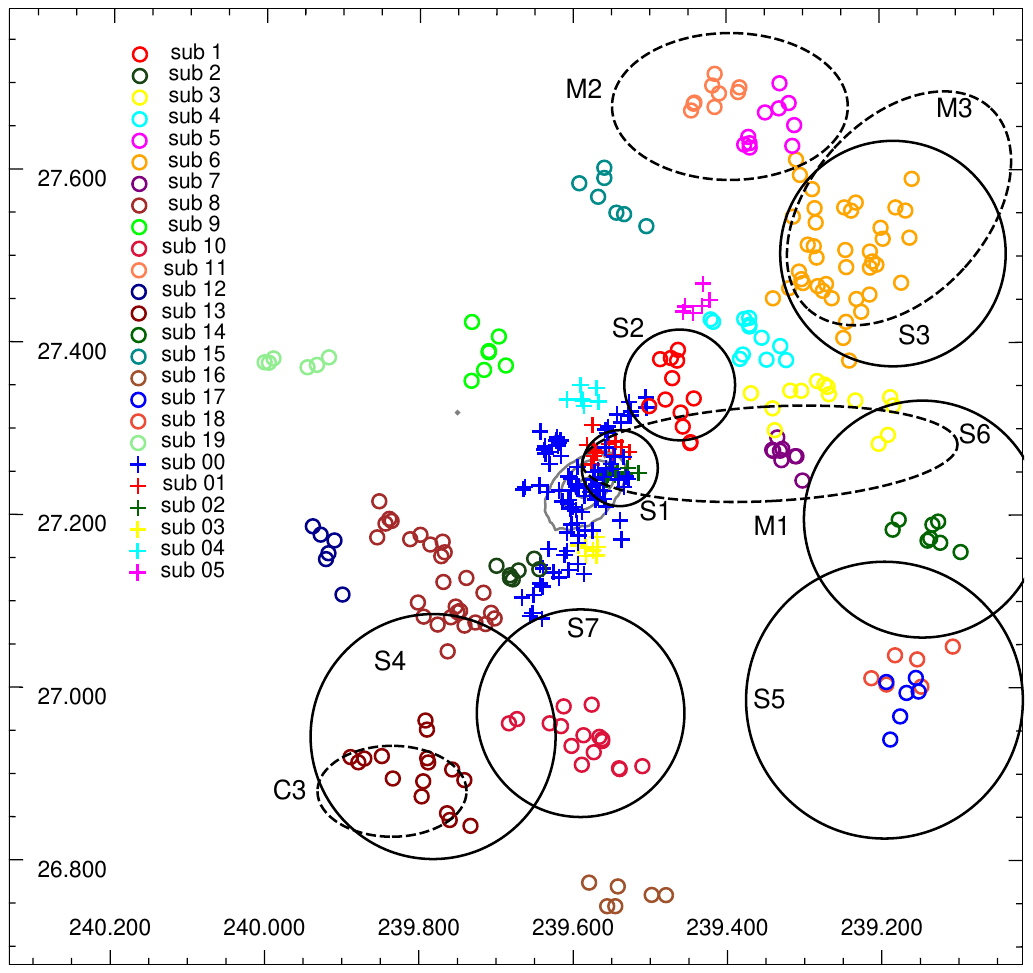}
\caption{Substructures identified with the $\sigma$-plateau algorithm and with other methods in the literature. S1--S7 shown as black circles show the location and size of the substructures identified with the $\kappa$-test by \citet{owers2011}. The dashed ellipses show the approximate location and size of the structures identified by \citet{einasto2017}. }
\label{compare}
\end{figure}

In this section, we combine the results of the above analyses to attempt to draw a scenario of the internal structure and dynamics of A2142. We also compare our results with the previous results by \citet{owers2011} and \citet{einasto2017}, who use different methods to detect substructures in A2142.

We first compare the optical substructures identified with the $\sigma$-plateau algorithm with the ICM redshift map, as shown in Figure \ref{redshift}. The  ICM redshift map only covers the
region within 3 arcmin from the cluster center, so Figure \ref{redshift} only shows the
substructures sub00 to sub03 identified with the third threshold trimming the binary tree of
A2142. Figure \ref{redshift} does not show any clear correlation between the redshift
and spatial distributions
of the optical substructures and the ICM distribution and redshift. This result suggests that the dynamics of the ICM in this region have decoupled from the dynamics of the galaxies. This behaviour is not unexpected in merging systems, because galaxies approximately behave like collisionless components, unlike the ICM.

In Figure \ref{wl}, we superimpose the substructures identified by the $\sigma$-plateau algorithm on the weak-lensing mass map. Figure \ref{wl} shows that the shape of the main halo of the mass map and the distribution of the member galaxies of the cluster core sub00 are consistent with each other.
Moreover, the $\sim3.3\sigma$ excess in the weak-lensing signal, located $\sim7'$--$8'$ northwest of the cluster center, coincides with sub1. This match supports the result of \citet{okabe2008}, who found that this northwest mass substructure is associated with a slight excess of galaxies in the color-magnitude relation of the A2142 galaxies, lying $\sim 5'$ ahead of the northwest edge of the central X-ray core: we confirm that 6 out of the 12 members of sub1 do indeed belong to this group of galaxies identified by \citet{okabe2008}. Finally, we also find marginally significant excesses in the weak-lensing map that coincide with sub04, sub01, and sub2. These results demonstrate that our $\sigma$-plateau substructure identification algorithm can efficiently detect structures that are around the detection limit of the weak-lensing signal.

Figure \ref{compare} compares the substructures identified with the $\sigma$-plateau algorithm with the substructures identified in previous work.
\citet{owers2011} detect 7 substructures, S1--S7, from the projected galaxy surface density distribution and a $\kappa$-test on the local kinematics of the galaxies, where the $\kappa$-test identifies kinematic substructures by comparing the local velocity distribution to the global velocity distribution \citep{colless1996}. Open circles in Figure \ref{compare} show the location and size of the substructures identified by \citet{owers2011}, according to their Figure 12.

\citet{einasto2017} identify four substructures, M1--M3 and C3, by analysing the position and velocities of member galaxies with the $mclust$ package, which is based on the analysis of a finite mixture of distributions, in which each mixture component corresponds to a different subgroup. To provide a qualitative impression of the location and size of these structures we plot four ellipses in Figure \ref{compare},
according to the information that can be inferred by eye from Figure 4 and 6 of \citet{einasto2017}.

We can see that sub1 overlaps S2 of \citet{owers2011}, which is consistent with the most
obvious DS substructure (see Figure \ref{fig:ds}). Sub5 and sub11 coincide with substructure M2 in \citet{einasto2017}.  Our most prominent substructure sub6 on the northwest coincides with S3 and M3; sub6 is also clearly detected by the DS analysis (Figure \ref{fig:ds}). Sub7 lies within M1; we note that the rest of the galaxies associated to M1 by \citet{einasto2017} are identified as substructure members by neither \citet{owers2011} nor our $\sigma$-plateau algorithm. Sub10 overlaps S7. Sub13 overlaps S4 and the component C3 of \citet{einasto2017} (see their figure 4). Sub17 and sub18 overlap S5 and also appear as clumps of the DS analysis.

All the remaining substructures, except sub8, that are identified by the $\sigma$-plateau algorithm and do not have a correspondence with substructures from previous analyses, have 14 members at most, indicating that either the catalogues used in previous analysis did not contain enough galaxies or these structures, if they are not chance alignment of unrelated galaxies, are too poor to be reliably identified by other methods. We conclude that, overall, this comparison shows a remarkably agreement between the different substructure identifications.

\begin{figure}
\centering
\includegraphics[width=0.49\textwidth]{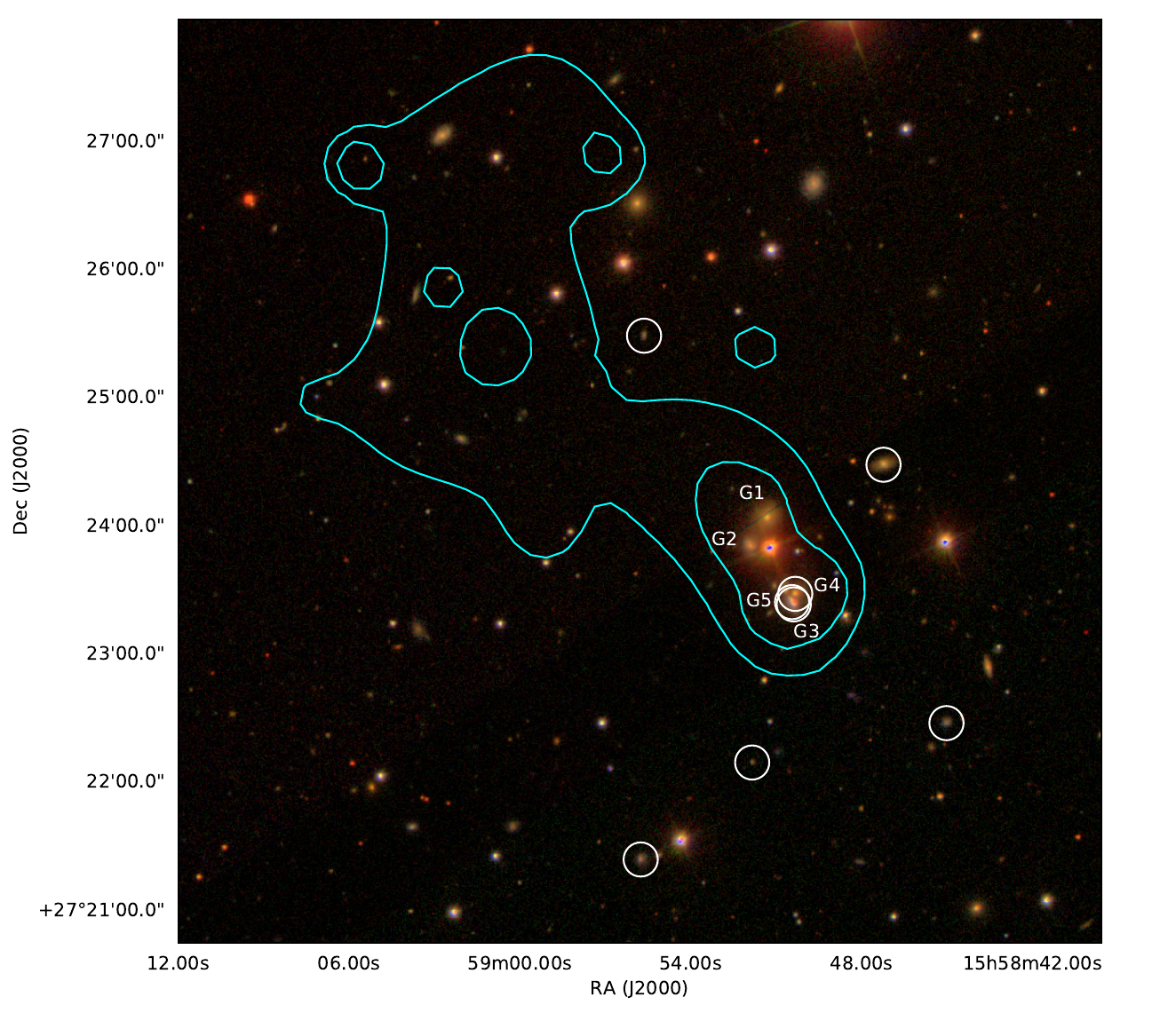}
\caption{The zoomed-in image of the substructure sub9. The SDSS RGB image converted from $i,r,g$-band is superimposed with the X-ray contours of the gas; the white circles are the members  of sub9 according to the $\sigma$-plateau algorithm. Five galaxies in the image are labeled G1--G5 as in \citet{eckert2017}. G3 and G4 are confirmed to be members of sub9.}
\label{infall}
\end{figure}

As shown in Sect.~\ref{sec4}, sub9 in the northeast of the cluster is identified by both the DS method and the $\sigma$-plateau algorithm. The circles associated to the galaxies according to the DS method in that region of the sky (Figure \ref{fig:ds}) have both red and blue colors: most of the red circles are members of sub9, whereas the blue circles, that have redshift smaller than the sub9 redshift, are not members of sub9. The X-ray images from both XMM-{\sl Newton} \citep{2014Eckert} and {\sl Chandra} \citep{eckert2017} show the presence of a faint gas component associated to this structure. For this X-ray emitting gas, \citet{2014Eckert, eckert2017} determine an average temperature of $\sim$1.4 keV, appropriate for a system with total mass of $\sim5.1\times10^{13}~M_{\odot}$ \citep{vikhlinin2006}. According to the analysis of \citet{munari2013,munari2014}, this mass is consistent with the mass
suggested by the velocity dispersion of 310 km$\rm\ s^{-1}$ that we measure for the members of the optical substructure. Figure \ref{infall} shows an optical
image of sub9, with the open white circles indicating its members according to the $\sigma$-plateau algorithm. We also indicate the five galaxies, G1--G5, associated by  \citet{2014Eckert,eckert2017} to the clump of hot gas.  Our analysis confirms that G3, that we name BG5 in section \ref{sec3.1}, G4 and G5 are members of sub9. On the contrary, G1 and G2 are not members: in fact, G1 and G2 have a velocity 1700~km~s$^{-1}$ smaller and 1130~km~s$^{-1}$ larger, respectively, than the mean redshift of sub9.
The redshift of sub9 -- $z=0.09459$  -- is  significantly larger than the average redshift of the cluster -- $z=0.08982$ -- and shows that this system is falling into the cluster at large speed.

We conclude that the optical substructure sub9 and the faint X-ray clump originate from the same group that is currently falling into the cluster: the ram pressure of the ICM acting on the group gas, but not on its galaxy members, is a plausible explanation for the displacement on the sky between the galaxies and the X-ray emission.

\section{Conclusions}
\label{sec7}
We investigate the dynamics of A2142 by comparing the properties of the
substructures in the galaxy distribution, the line-of-sight velocity field of the
ICM derived from the spatially-resolved X-ray spectroscopy, and the weak-lensing mass distribution. Our main results are as follows:

\begin{itemize}

 \item Based on a new and extended catalog of spectroscopic redshifts within $\sim 3.5$~Mpc from the cluster center, we identify a number of substructures with the $\sigma$-plateau algorithm.
The distribution of the substructures appears consistent with results obtained with other methods, including the DS method, the $\kappa$-test and the $mclust$ algorithm.

 \item Most substructures have a number of galaxy members $\sim 10$ (see Table \ref{table:subs}), indicating that there is no sign of recent or ongoing major mergers and suggesting a scenario where the cold fronts observed in A2142 in X-rays originate from core sloshing induced by minor mergers.

 \item In the northeast outskirts, the galaxy substructure sub9 matches a falling gas clump observed in X-rays; the slight displacement between the positions of the galaxies and the gas might be due to ram pressure on the hot gas.

 \item The shape of the central substructure sub00 is consistent with the projected weak lensing mass map.
 Several substructures also coincide with the weak-lensing mass excesses.

 \item With spatially-resolved X-ray spectroscopy based on {\sl Chandra} data, we measure the line-of-sight velocity distribution of the ICM within $\sim 0.35$~Mpc from the cluster center. We find an annulus near the X-ray cold fronts with redshift significantly larger than the surroundings, corresponding to a velocity $810\pm330 {\rm km~s^{-1}}$ larger than the cluster mean velocity. We also find that the core is surrounded by high redshift gas, with a velocity $660\pm300 {\rm km~s^{-1}}$ larger than the cluster redshift. The features we observe in the X-ray redshift map appear to be consistent with the core-sloshing scenario suggested in previous work.

\end{itemize}

Deeper photometric and spectroscopic observations of the
field can clearly provide more detailed  and solid results.
The spatially resolved X-ray redshift measurements will
improve by advanced future  X-ray bolometers. In
particular, with the X-ray IFU on board, the Advanced Telescope
for High-ENergy Astrophysics ({\sl Athena}) will remarkably
extend the application of this method. Moreover, a more
precise weak lensing measurement of the projected mass distribution
will be of great help to confirm the relation between
the total mass distribution and the galaxy substructures we
find here.

\acknowledgments
We thank the anonymous referee whose comments substantially improved the presentation of our results.
We thank Dominique Eckert, Sabrina de Grandi, and Maria Chiara Rossetti for the results of XMM-{\sl Newton} and their helpful comments. We sincerely
thank Margaret Geller for her insightful suggestions.
This work was supported by the National Natural Science Foundation of China under
Grants Nos. 11403002, the Fundamental Research Funds for the Central Universities
and Scientific Research Foundation of Beijing Normal University.
AD and HY acknowledge partial support from the grant of the Italian Ministry of Education, University and Research (MIUR) (L. 232/2016) ``ECCELLENZA1822$\_$D206 - Dipartimento di Eccellenza 2018-2022 Fisica'' awarded to the Dept. of Physics of the University of Torino.
AD also acknowledges partial support from the INFN Grant InDark.
P.T. is supported by the Recruitment Program of High-end Foreign Experts
and he gratefully acknowledges hospitality of Beijing Normal University. K.U. acknowledges support from the Academia Sinica Investigator Award
and from the Ministry of Science and Technology of Taiwan (grant MOST
103- 2628-M-001-003-MY3).

\bibliography{a2142}
\end{CJK*}
\end{document}